\documentclass[preprint]{elsarticle}
\usepackage{amssymb}
\usepackage{amsthm}
\usepackage[T1]{fontenc}
\usepackage{mathtools}
\usepackage{breqn}
\usepackage{xfrac}
\usepackage{braket}
\usepackage{physics}
\usepackage{tikz}
\usetikzlibrary{automata,er,positioning,bayesnet}
\usepackage{caption}
\usepackage{subcaption}
\usepackage{xparse}
\usepackage{float}
\usepackage{varwidth}
\usepackage[inline]{enumitem}
\usepackage{xcolor}
\usepackage[noabbrev,capitalize]{cleveref}
\usepackage[normalem]{ulem}

\journal{Theoretical Computer Science}

\usetikzlibrary{fit, shapes.geometric, fadings, patterns, positioning, quotes}

\tikzfading[name=fade out, inner color=transparent!0, outer color=transparent!100]

\usepackage[pagewise]{lineno}

\newtheorem{theorem}{Theorem}
\newtheorem{lemma}[theorem]{Lemma}
\newtheorem{corollary}[theorem]{Corollary}

\NewDocumentEnvironment{delineate}{m}{\textcolor{cyan!70!black!}{> > > > Begin: #1 > > > >}}{\textcolor{red!70!black!}{< < < < End: #1 < < < <}}

\mathchardef\mhyphen="2D

\DeclarePairedDelimiter{\ceil}{\lceil}{\rceil}

\newcommand{\lend}{\ensuremath{\rhd}}

\def\runo #1#2{#1_1, #1_2, ..., #1_{#2}}

\def\codinglength {\sum_{i=1}^n\log_2(\chi(u_i,\sigma_i))}

\def\codinglengthpsi {\sum_{i=1}^n\log_2(\psi(u_i,\sigma_i))}

\begin{document}

\begin{frontmatter}

\title{Energy Complexity of Regular Languages\footnote{A preliminary version of this paper appeared as \cite{YKUS22}.}}

%

\author{$\text{Fırat Kıyak}\footnote{Department of Mathematics, Boğaziçi University, İstanbul, Turkey}$,   $\text{A. C. Cem Say}\footnote{Department of Computer Engineering, Boğaziçi University, İstanbul, Turkey}$}
%

%
%

%
%
%
%

\begin{abstract}
Each step that results in a bit of information being ``forgotten'' by a computing device has an intrinsic energy cost. Although any Turing machine can be rewritten to be thermodynamically reversible without changing the recognized language, finite automata that are restricted to scan their input once in ``real-time'' fashion can only recognize the members of a proper subset of the class of regular languages in this reversible manner. We study the energy expenditure associated with the computations of  deterministic and quantum finite automata.  We prove that zero-error quantum finite automata have no advantage over their classical deterministic counterparts in terms of the maximum obligatory thermodynamic cost  associated by any step during the recognition of different regular languages. We also demonstrate languages for which ``error can be traded for energy'', i.e. whose zero-error recognition is associated with computation steps having provably bigger obligatory energy cost when compared to their bounded-error recognition by real-time finite-memory quantum devices.  We show that regular languages can be classified according to the intrinsic energy requirements on the recognizing automaton as a function of  input length, and prove upper and lower bounds.
\end{abstract}


\end{frontmatter}


\section{Introduction}

The discovery of the relationship between thermodynamics and computation, revealing the links between the concepts of heat, entropy, and information, is a landmark scientific achievement \cite{LR02}. As shown by Landauer \cite{Lan61}, each bit of information ``forgotten'' by a computing device is associated with the dissipation of an amount of heat proportional to the absolute temperature of the device, and thus an unavoidable minimum energy cost for any fixed temperature. Turing machine programs \cite{Ben73} (and even finite automata with two-way access to their input strings \cite{KW97}) can be rewritten to be \textit{reversible}, so that each one of their configurations has a single possible predecessor, and  their computational steps can therefore in principle be executed using arbitrarily small amounts of energy, but things change when one limits attention to real-time finite automata.

It is known \cite{Pin92} that reversible real-time finite automata (where each state has at most one incoming transition with each possible symbol of the input alphabet) recognize only a proper subset of the class of regular languages, so some regular languages necessarily have automata with states receiving multiple transitions with the same symbol. Intuitively, it is impossible to ``rewind''  computations of such machines, since they ``forget'' which one of a set of possible predecessor states led them to their present state. It is natural to ask if these ``in-degrees'' can be used to classify regular languages in terms of the minimum energy required for their recognition by such automata. 

It was precisely because of the reversibility requirement inherent in unitary matrices that early definitions of real-time quantum finite automata (QFAs) \cite{MC00,BC01,KW97} were not able to capture all regular languages. Modern definitions of QFAs \cite{Hir10,SY14a}, which recognize all and only the regular languages with bounded error, are able to handle irreversible languages by using not one but many instances of an architectural component (called an \textit{operation element}) that can be seen to correspond to the previously mentioned notion of ``incoming transitions'' in deterministic finite automata (DFAs), so the energy question raised above is relevant for the study of bounded-error QFAs as well.

Energy is a  resource (like time and memory space) that has been studied  \cite{LV96,LMT00} in the context of computational complexity theory. This paper takes up this study in the additionally restricted framework of finite automata. 
We start (Section \ref{sec:genQFA}) by clarifying the intuitive notion that links the number of incoming transitions received by a state to energy expenditure by using the general QFA definition of \cite{SY14a}  to model the information loss inherent in the computations of such machines, and using Landauer's principle. We study the problem of determining the maximum obligatory thermodynamic cost of any step associated with the recognition of a given regular language by a finite automaton in the following three sections.
In Section \ref{sec:zeroerror}, we show that zero-error quantum finite automata have no advantage regarding this energy cost  over their classical deterministic counterparts, which are restricted to contain just 0's and 1's in their transition matrices.
  Section \ref{sec:hier} establishes 
an upper bound on the number of bits that have to be ``forgotten'' during any step of a computation associated with the recognition of any regular language, namely, any such language on an alphabet with $k$ symbols can be recognized by a  DFA with the property that no  state  has more than $k+1$ incoming transitions labeled with the same input symbol, and thus requires no more than $\log_2(k+1)$ bits to be erased by any step. 
In Section \ref{sec:boundederror}, we  demonstrate languages for which ``error can be traded for energy'', i.e. whose zero-error recognition is associated with provably bigger energy cost for some steps when compared to their bounded-error recognition by real-time finite-memory quantum devices. 
Section \ref{sec:langcost} 
establishes lower and upper bounds on the energy complexities (defined as the number of forgotten bits as a function of input length) of irreversible languages, and demonstrates  that these languages can be classified in terms of their energy complexities.
Section \ref{sec:conc} is a conclusion.
   



\section{Information erasure by finite automata and the general QFA model}\label{sec:genQFA}

It is important to start by clarifying the nature of the resource whose usage is studied in this paper.  A computing device is said to ``forget'' $\log_2 P$ bits of information when it arrives at a configuration which has $P$ predecessors, this increases the entropy by $k_B \ln{2}\log_2 P$ joules per kelvin (where $k_B$ is Boltzmann's constant), and the amount of dissipated heat can be calculated by multiplying this entropy difference by the ambient temperature $T$ \cite{Lan61}. It is evident that the energy cost under consideration is directly proportional to the number of forgotten bits. We will therefore define the \textit{energy complexity}  of a finite automaton as the maximum number of bits of information that it forgets while running  on any input string of a given length. Although we considered the name  ``entropy complexity'', we decided that  ``energy'' is a much more natural concept to associate with the notion of a ```resource'', when compared to entropy.\footnote{The term ``information complexity'', which has already been appropriated for other purposes, would also be unsuitable, since the information that our machines forget is not ``possessed'' by them before the start of the computation, as will be seen shortly.}

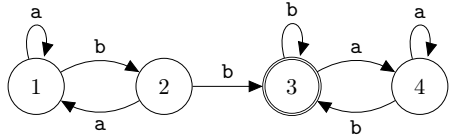
\begin{figure}
        \centering
    \scalebox{0.85}{
        \begin{tikzpicture}
        
             \node[state] (one) {1};
             \node[state, xshift=2cm] (two) {2};
             \node[state, accepting, xshift=4cm] (three) {3};
             \node[state, xshift=6cm] (four) {4};
             
            \draw   (one) edge[bend left, above, ->] node{\texttt{b}} (two)
                    (two) edge[bend left, below, ->] node{\texttt{a}} (one)
                    (one) edge[loop above] node{\texttt{a}} (one)
                    (two) edge[above, ->] node{\texttt{b}} (three)
                    (three) edge[loop above] node{\texttt{b}} (three)
                    (three) edge[bend left, above, ->] node{\texttt{a}} (four)
                    (four) edge[bend left, below, ->] node{\texttt{b}} (three)
                    (four) edge[loop above] node{\texttt{a}} (four);
        \end{tikzpicture}
        }
    \caption{A DFA transition diagram}
    \label{fig:f1}
     \end{figure}

Consider the DFA transition diagram in Fig. \ref{fig:f1}. If the machine is known to have arrived in state 3 by consuming the symbol \texttt{b} from the input, there is no way to tell which of the three states in $\{2,3,4\}$ it was at in the previous step. One sees that state 3 is associated with the maximum energy cost that can be incurred by individual steps during computations of this machine, whereas state 2 incurs no such cost. 

In the following three sections, we will focus on quantifying this maximum energy cost that can be incurred by any particular step of an automaton's execution. It will be instructive to model this process by using a quantum finite automaton model, which clearly differentiates the execution stages associated with energy expenditure  from the other stages that are carried out by unitary (and therefore reversible) transformations.

Although classical physics, on which the intuition underlying deterministic computation models is based, is supposed to be subsumed by quantum physics, early definitions of quantum finite automata (e.g. \cite{MC00,KW97}) resulted in ``weak'' machines that could only capture  a proper subset of the class of regular languages. The cause of this apparent contradiction was identified \cite{Hir10} to be those early definitions' imposition of unnecessarily strict limitations on the interaction of the automata with their environments. Classical finite automata, after all, are not ``isolated'' systems, with loss of information about the preceding configuration and the ensuing transfer of heat to the environment implied by their logical structure. The modern definition of QFAs, \cite{Hir10,SY14a} which we use below to illustrate   the information loss associated with the ``costliest'' computation step, allows a sufficiently rich repertory of unitary transformations and measurements that does not overrule any physically realizable finite-memory computation.\footnote{References \cite{AY18} and \cite{SY14a} provide a more comprehensive introduction to the quantum computation notation and concepts discussed here.}

A \textit{quantum finite automaton} (QFA) is a 5-tuple $(Q,\Sigma,\{\mathcal{E}_\sigma | \sigma \in \Sigma_\lend \},q_1,F)$, where
\begin{enumerate}
    \item $Q=\{q_1,\ldots,q_n\}$ is the finite set of machine states,
    \item $\Sigma$ is the finite  input alphabet,
    \item $q_1 \in Q$ is the  initial state,
    \item   $F \subseteq Q$ is the set of accepting states, and
    \item  $\Sigma_\lend = \Sigma \cup \Set{\lend}$, where $\lend \notin \Sigma$ is  the left  end-marker symbol, placed automatically before the input string, and for each $\sigma \in \Sigma_\lend$, $\mathcal{E}_\sigma$ is the superoperator describing the transformation on the current configuration of the machine associated with the consumption of the symbol $\sigma$. For some $l \geq 1$, each $\mathcal{E}_\sigma$ consists of  $l$ operation elements $\{E_{\sigma,1},\ldots,E_{\sigma,l}\}$, where each operation element is a complex-valued $n \times n$ matrix.  
\end{enumerate}

Although it is customary in the literature to analyze these machines using density matrices \cite{Hir10,AY18}, we  take the alternative (but equivalent) approach of  \cite{SY14a}, which
represents the ``periphery'' that will support intermediate measurements during the execution of our QFA by an \textit{auxiliary  system} with the state set $\Omega = \{\omega_1, ..., \omega_l\}$.
The thermodynamic cost of computational steps is made explicit by considering an additional set $S=\{s_1, ..., s_l\}$ of classical states, which represent the information to be forgotten by the machine. 
These  states  will mirror the members of $\Omega$ during computation, as will be described below. 

Considered together, the auxiliary system and the ``main system'' of our machine defined above have the state set $\Omega \times Q$. The quantum state space of the overall system is $\mathcal{H}_l \otimes \mathcal{H}_n$, the composite of the corresponding finite-dimensional Hilbert spaces.
Initially, this composite system is in the quantum state $\ket{\omega_1}\otimes\ket{q_1}$, and the classical state is $s_1$. At the beginning of every computational step, it will be ensured that the auxiliary system is again at one of its computational basis states, i.e. $\ket{\omega_j}$ for some $j$, and the classical state will be $s_j$.

Let $\ket{\psi_{x}}= \alpha_1\ket{q_1}+....+ \alpha_n\ket{q_n}$ denote any vector in $\mathcal{H}_n$ that is attained by our QFA with nonzero probability after it has consumed the string $x \in \Sigma^*$.   
We will examine the evolution of the overall system for a single step starting at  a state $\ket{\omega_j}\otimes\ket{\psi_x}$. If the symbol $\sigma$ is consumed from the input, the composed system first undergoes the unitary operation described by the product 
$U_{\sigma} U_{s_j}$, as described below.

$U_{s_j}$ is designed so that its application  rotates the auxiliary state from $\omega_j$ to $\omega_1$, so that $U_{\sigma}$ will act on
\[
    \ket{\Psi_x} =
     \ket{\omega_1}\otimes \ket{\psi_{x}}
    =
    (\underbrace{\alpha_1, \alpha_2, ..., \alpha_n}_{\text{amplitudes of }\ket{\psi_x}}, \underbrace{0, 0, ..., 0}_{(l-1)\times n \text{ times}})^T.
    \]
 
Only the leftmost $n$ columns of the matrix $U_{\sigma}$ are significant for our purposes, and the remaining ones can be filled in to ensure unitarity. Those first $n$ columns will be provided by the operation elements $E_{\sigma,1},\ldots,E_{\sigma,l}$, as indicated by the following partitioning of $U_{\sigma}$ into $n \times n$ blocks:
\[
U_{\sigma} = 
    \begin{bmatrix}
    \begin{array}{c|c|c|c}
      E_{\sigma,1} & * & ... & * \\
      \hline
      E_{\sigma,2} & * & ... & * \\
      \hline
      \vdots & \vdots & \ddots & \vdots \\ 
      \hline
      E_{\sigma,l} & * & ... & *
    \end{array}
    \end{bmatrix}
\]

(Since $U_{\sigma}$ is unitary, one sees that the operation elements should satisfy the constraint
    $\sum_{j=1}^{l} E_{\sigma,j}^\dag E_{\sigma,j} = I$.)

Consider the $n$-dimensional vectors $\widetilde{\ket{\psi_{x\sigma,i}}}=E_{\sigma,i}\ket{\psi_x}$ for $i \in \{1,\ldots,l\}$. Clearly, the vector $\widetilde{\ket{\Psi_{x\sigma}}}=U_{\sigma}\ket{\Psi_x}$ that represents the overall system state obtained after the unitary transformation described above can be written by ``stacking'' these vectors, each of which corresponds to a different auxiliary state,  on top of each other, as seen in Equation \ref{eq:longvector}. 

\begin{equation}\label{eq:longvector}
    \widetilde{\ket{\Psi_{x\sigma}}} 
=
\begin{bmatrix}
\widetilde{\ket{\psi_{x\sigma,1}}} \\ 
\widetilde{\ket{\psi_{x\sigma,2}}} \\ 
\vdots \\
\widetilde{\ket{\psi_{x\sigma,l}}}
\end{bmatrix}
= \ket{\omega_1}\otimes \widetilde{\ket{\psi_{x\sigma,1}}}+ \ket{\omega_2}\otimes \widetilde{\ket{\psi_{x\sigma,2}}} + \ldots + \ket{\omega_l}\otimes \widetilde{\ket{\psi_{x\sigma,l}}}
\end{equation}

At this point in the execution of our QFA, the auxiliary system is measured in its computational basis. The probability $p_k$ of observing outcome ``$\omega_k$'' out of the $l$ different possibilities is the square of the length of $\widetilde{\ket{\psi_{x\sigma,k}}}$. As a result of this probabilistic branching, the main system collapses to the state $\ket{\psi_{x\sigma,k}} = \frac{\widetilde{\ket{\psi_{x\sigma,k}}}}{\sqrt{p_k}}$ with probability $p_k$ (for $k$ such that $p_k>0$), and the fact that this observation result is recorded for usage in the next step is represented by setting  the classical state to $s_k$, overwriting its previous value.\footnote{Each observation with $l>1$ possible outcomes would necessarily leave the observer in one of $l$ different states. A finite-state observer can store only a fixed number of past observation results in memory, and cannot perform an arbitrarily long sequence of observations without overwriting (forgetting) some of the older outcomes. We model the case where this overwriting happens at every step.
} 
 It is only this final action of ``forgetting'' the previous 
 value of the classical state that can cause an energy cost for a step of a QFA:  This information amounts to at most $\log_{2}l$  bits, and one needs to expend a minimum of $k_B T \ln{2}$ joules to erase each bit, where $T$ is the ambient temperature in kelvins, as described above \cite{Lan61}. 
(We note again that this value is the obligatory energy requirement of not all, but only the most ``expensive'' steps during the recognition of the language. The QFA model we use forces all transitions related to all input symbols to be represented by the same number ($l$) of operation elements. We will be using a more flexible approach to take the different, possibly lower,  individual energy costs of different transitions in Section \ref{sec:langcost}. The present framework will be useful for clarifying the relationship between the DFA and QFA models, and for studying energy costs of genuinely quantum machines in the next section.)

After processing the entire input string symbol by symbol in the manner described above, the main system, described by some $n$-dimensional vector $\ket{\psi}$, is measured in its computational basis. The probability of acceptance at this point is the sum of the squares of the lengths of the amplitudes of the accepting states in $\ket{\psi}$. Rejection is similarly defined in terms of the non-accepting states. A language $L$ is said to be recognized by a QFA with \textit{bounded error} if there exists a number $\epsilon < \frac{1}{2}$ such that every string in $L$ is accepted and every string not in $L$ is rejected by that QFA with probability at least $1-\epsilon$. If $\epsilon=0$, i.e. the QFA has the property that it accepts every input string with either probability 0 or 1, it is said to recognize the set of strings that it does accept with \textit{zero error}.

It is known \cite{LQZLWM12} that ``modern'' QFAs defined in this manner can recognize all and only the regular languages with bounded error.\footnote{``Zero-energy'' QFAs with a single operation element in each superoperator correspond to the earliest definition in \cite{MC00,BC01}, and can  recognize all and only the group languages (a proper subclass of the class of regular languages, whose DFAs have the property that one again obtains a DFA if one reverses all their transitions) with bounded error \cite{BC01,BP02}.} Given any DFA with $n$ states, it is straightforward to build a QFA with $n$ machine states that recognizes the same language $M$ with zero error. An examination of this construction is useful for understanding the nature of the information lost when the classical state is overwritten during a computational step of a QFA.

 \begin{figure}[htp]
 \centering
\begin{subfigure}{1\textwidth}
        \centering
    \scalebox{0.85}{
        \begin{tikzpicture}
        
             \node[state, initial] (one) {1};
             \node[state, xshift=2cm] (two) {2};
             \node[state, accepting, xshift=4cm] (three) {3};
             \node[state, xshift=6cm] (four) {4};
             
            \draw   (one) edge[bend left, above, ->] node{\texttt{b}} (two)
                    (two) edge[bend left, below, ->] node{\texttt{a}} (one)
                    (one) edge[loop above] node{\texttt{a}} (one)
                    (two) edge[above, ->] node{\texttt{b}} (three)
                    (three) edge[loop above] node{\texttt{b}} (three)
                    (three) edge[bend left, above, ->] node{\texttt{a}} (four)
                    (four) edge[bend left, below, ->] node{\texttt{b}} (three)
                    (four) edge[loop above] node{\texttt{a}} (four);
        \end{tikzpicture}
        }
    \caption{DFA transition diagram}
    \label{fig:sub1}
     \end{subfigure}
     \newline
     \newline
     \newline
     \begin{subfigure}{0.3\textwidth}
     \centering
        \includegraphics[width=0.8\textwidth]{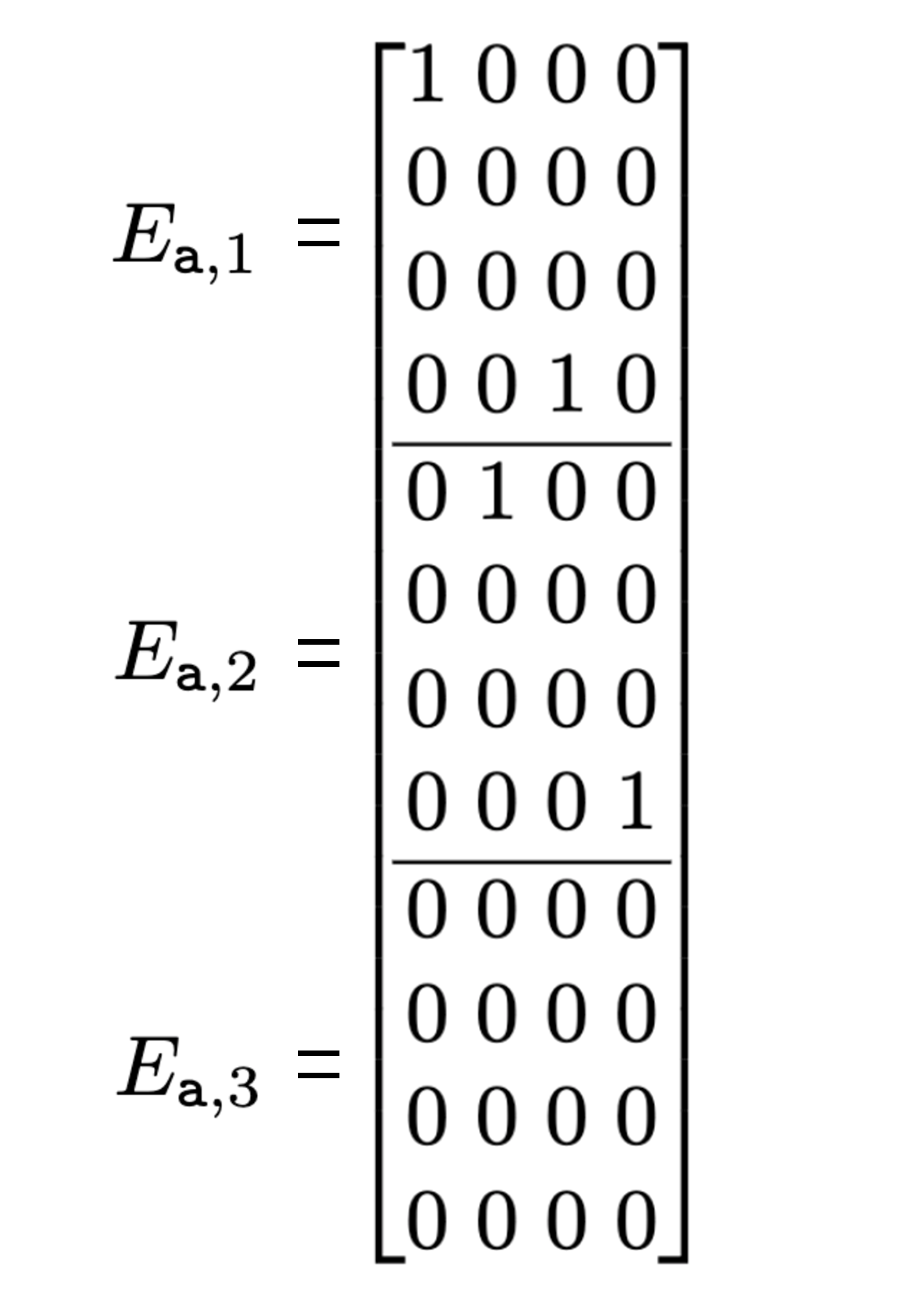}
        \caption{Superoperator for $\texttt{a}$}
        \label{fig:sub2}
     \end{subfigure}
     \begin{subfigure}{0.3\textwidth}
     \centering
        \includegraphics[width=0.8\textwidth]{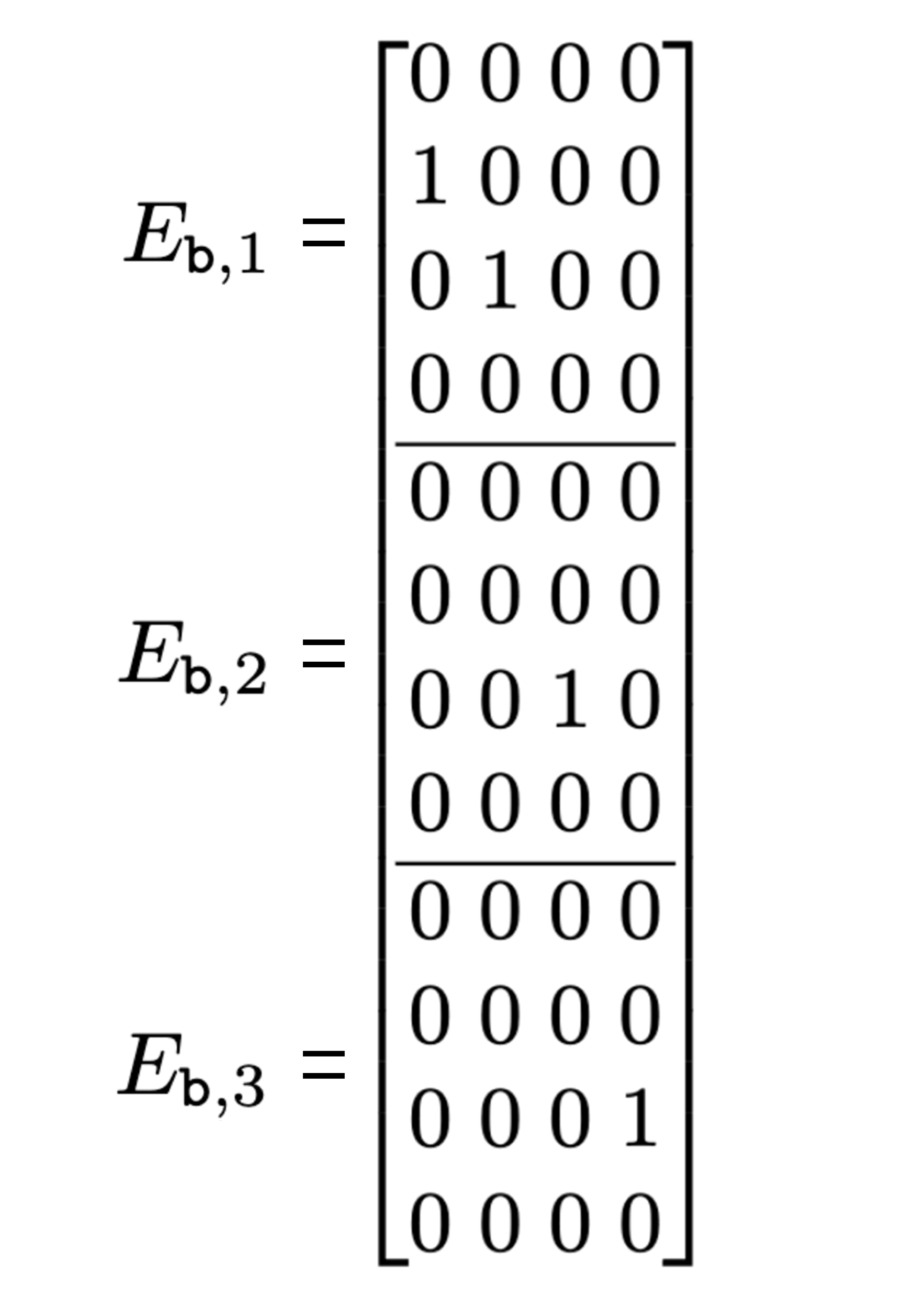}
        \caption{Superoperator for $\texttt{b}$}
        \label{fig:sub3}
     \end{subfigure}
     \caption{A DFA and superoperators for its QFA implementation}
     \label{fig:QFAmatrix}
 \end{figure}

Consider the DFA of Fig. 1, whose transition diagram is replicated in Fig. \ref{fig:sub1}.   Figures \ref{fig:sub2} and \ref{fig:sub3} depict the operation elements associated with  symbols $\texttt{a}$ and $\texttt{b}$ in the QFA implementation of this machine.\footnote{The left end-marker is inconsequential in DFA simulations, and its superoperator is not shown.} In each square matrix, both the rows and columns correspond to the states of the QFA, which in turn correspond to the states of the DFA of  Fig. \ref{fig:sub1}. The entry at row $i$, column $j$ of the $k$'th operation element for a symbol represents the  transition that the QFA would perform  from its $j$'th machine state to the combination of  its $i$'th machine state and $k$'th auxiliary state upon consuming that symbol. Starting with the vector $(1, 0, 0, 0)^T$ representing the machine being at the initial state with probability 1, the QFA would trace every step in the execution of the DFA on any input string, and recognize the same language with zero error. 
    %
             


The reader will note in Fig. \ref{fig:QFAmatrix} that the superoperators, which are just adjacency matrices for the DFA,  have not one, but three operation elements precisely because state 3 has three incoming transitions labeled with the same symbol in Fig. \ref{fig:sub1}: We cannot have two 1's in the same row of two different columns of the matrices in Figures \ref{fig:sub2} and \ref{fig:sub3}, since they must be orthonormal. We use the additional operation elements to represent the additional ways in which the machine can switch to state 3 with input $\texttt{b}$. Intuitively, the auxiliary state records which of the three transitions was used to enter state 3, and it is not possible to ``trace the computation backwards'' from that state when one has forgotten that information.\footnote{Since none of the three states with $\texttt{b}$-transitions into state 3 is more likely to be the source than the others, this information amounts to $\log_2 3$ bits.} This is why the language recognized by these machines is not ``reversible'' \cite{Kut14}.

We have seen how any DFA with $n$ states and at most $l$ incoming transitions to the same state with the same symbol can be simulated by a zero-error QFA with $n$ machine states and $l$ operation elements (that only contain 0's and 1's) per superoperator. Note that the QFAs that are constructed to imitate DFAs in the fashion exemplified above do not use any ``quantumness'': At all times, the state vector of the QFA never represents a superposition of more than one classical state, and just tracks the execution of the DFA faithfully. There is no probabilistic ``branching'' (since only one auxiliary state has nonzero amplitude at any step), and no constructive or destructive interference among amplitudes. It is natural to ask if QFAs with other complex-valued entries in their operation elements can utilize the famous non-classical capabilities of quantum systems to perform the same task employing more energy-efficient steps, i.e. with fewer operation elements. We answer this question in Section \ref{sec:zeroerror}.

\section{Zero-error QFAs have no energy advantage}\label{sec:zeroerror}

For any language $L$ defined over alphabet $\Sigma$,  the ``indistinguishability'' relation $\equiv _{L}$ on the set $\Sigma^*$ is defined as follows: 
\[(x \equiv _{L} y) \iff (\forall z \in \Sigma ^* [xz \in L \iff yz \in L]).
\]
\begin{lemma}
Let  $M$ be a QFA recognizing a language $L$ with zero error. Let $x$ and $y$ be strings such that $x \not\equiv _{L} y$. If  $\ket{\psi_{x}}, \ket{\psi_{y}}\in \mathcal{H}_n$ are any two vectors  that are attained by $M$ with nonzero probability after it reads $x$ or $y$, respectively, then  $\braket{\psi_{x}}{\psi_{y}}=0$.
\end{lemma}
\begin{proof}
 Let us say that \textit{$x$ and $y$ are distinguishable with respect to $L$ in $k$ steps} if there exists a string $z$ of length $k$ that distinguishes them, i.e.  $xz \in L$ if and only if $yz \notin L$. We will prove the statement by  induction on  the number of steps in which $x$ and $y$ are distinguishable with respect to $L$.

\textbf{Basis:}

If $x$ and $y$ are distinguishable with respect to $L$ in 0 steps, let us say without loss of generality that  $x \in L$ and $y \notin L$. In this case, all entries of $\ket{\psi_{x}}$ corresponding to the non-accepting states in $Q$ must be zero, since $M$ would otherwise reject $x$ with nonzero probability. Similarly, all entries of $\ket{\psi_{y}}$ corresponding to the accepting states in $Q$ must be zero. But this means that  $\braket{\psi_{x}}{\psi_{y}}=0$.

\textbf{Induction step:}

Assume that the statement is true for all pairs of strings  that are distinguishable with respect to $L$ in $k$ steps, where $k\geq 0$, and consider the case of any $x$ and $y$ that are distinguishable  in  $k+1$ steps. In this context, we will further assume that  $\braket{\psi_{x}}{\psi_{y}}\neq 0$, and reach a contradiction.

Let $\sigma$ be the leftmost symbol of the  string $z$ (of length $k+1$) that distinguishes $x$ and $y$. Consider two copies of $M$ at states $\ket{\psi_{x}}$ and $\ket{\psi_{y}}$. When these two machines consume the input symbol $\sigma$, the corresponding vectors representing the composite system of the machine and its environment are both multiplied by the unitary matrix $U_\sigma$ to yield two $nl$-dimensional vectors, say,

\begin{equation}\label{eq:twoPsis}
    \widetilde{\ket{\Psi_{x\sigma}}} = 
\begin{bmatrix}
\widetilde{\ket{\psi_{x\sigma,1}}} \\ 
\widetilde{\ket{\psi_{x\sigma,2}}} \\ 
\vdots \\
\widetilde{\ket{\psi_{x\sigma,l}}}
\end{bmatrix}
\text{    and   }
\widetilde{\ket{\Psi_{y\sigma}}} = 
\begin{bmatrix}
\widetilde{\ket{\psi_{y\sigma,1}}} \\ 
\widetilde{\ket{\psi_{y\sigma,2}}} \\ 
\vdots \\
\widetilde{\ket{\psi_{y\sigma,l}}}
\end{bmatrix},
\end{equation}
where $n$ and $l$ are respectively the numbers of machine and auxiliary states, as we saw in Equation \ref{eq:longvector}. Since $U_\sigma$ preserves inner products and angles, these ``tall'' vectors are also not orthogonal by our assumption that $\braket{\psi_{x}}{\psi_{y}}\neq 0$.

As discussed in Section \ref{sec:genQFA}, the state vectors that $M$ can attain with nonzero probability after consuming this $\sigma$ are the normalized versions of the (nonzero) $n$-dimensional ``slices'' of $\widetilde{\ket{\Psi_{x\sigma}}}$ and $\widetilde{\ket{\Psi_{y\sigma}}}$. Note in Equation \ref{eq:twoPsis} that if $\braket{\widetilde{\psi_{x\sigma,j}}}{\widetilde{\psi_{y\sigma,j}}}= 0$ for all $j \in \{1,\ldots,l\}$, then  $\widetilde{\ket{\Psi_{x\sigma}}}$ and $\widetilde{\ket{\Psi_{y\sigma}}}$ must also be orthogonal. This means that $\braket{\psi_{x\sigma,j}}{\psi_{y\sigma,j}}\neq 0$ for at least one $j$, which is a contradiction, since $x\sigma$ and $y\sigma$ are distinguishable in $k$ steps.
\end{proof}
It follows that the subspace generated by the vectors attainable by $M$ through reading strings in a particular equivalence class of $\equiv _{L}$ must be orthogonal to all the subspaces corresponding to the other classes. 
This provides a new proof of the (already known) fact that zero-error QFAs can only recognize regular languages.
\begin{corollary}\label{cor:subspaces}
If a language $L$ is recognized by a
zero-error QFA $M$ with $n$ machine states, each equivalence class $C$ of $\equiv _{L}$ corresponds to a subspace $S_C$ of $\mathcal{H}_n$, and any two subspaces corresponding to different classes are orthogonal to each other. Since the sum of the dimensions of these subspaces is at most $n$,  $\equiv _{L}$ can have at most $n$ equivalence classes, and $L$ is regular by the Myhill-Nerode theorem.
\end{corollary}

We can now demonstrate that every zero-error QFA $M$ has a corresponding DFA $M'$ which recognizes the same language, and is as efficient as $M$ in terms of both memory (number of states) and the energy requirement of the costliest computational steps.\footnote{The fact that zero-error QFAs have no advantage over equivalent  DFAs in terms of the number of machine states was first proven by Klauck \cite{Klauck00}, using Holevo's theorem and communication complexity arguments.}
\begin{theorem}\label{th:qfatodfa}
For any $n,l>0$, if a language is recognized with zero error by a  QFA with $n$ machine states and $l$ operation elements per superoperator, then the same language is also recognized by a DFA with $n$ states and at most $l$ incoming transitions to the same state with the same symbol.
\end{theorem}
\begin{proof}
Let $M$ be a zero-error QFA with $n$ machine states and $l$ operation elements per superoperator. By Corollary \ref{cor:subspaces}, $M$ recognizes a regular language $L$. Let $k$ be the number of states of the minimal DFA $D$ recognizing $L$. Each input string $x$ that carries  $D$ to state $i\in\{1,\ldots,k\}$ of $D$ will carry $M$ to a state vector in a corresponding subspace $S_i$ of $\mathcal{H}_n$.\footnote{At this point, one may be tempted to declare that the set of subspaces already provides the state set of the DFA we wish to construct. After all, each matrix of the form $U_\sigma$ that we saw in Section \ref{sec:genQFA} ``maps'' a vector in $S_i$ to one or more vectors in $S_j$ if and only if $D$ switches from state $i$ to state $j$ upon consuming $\sigma$. However, this simple construction does not guarantee our aim of keeping the maximum number of incoming transitions with the same label to any state in the machine to a minimum.} 
Consider the DFA $M'$ that is 
described by the 5-tuple
 $(Q,\Sigma,\delta,q_1,F)$, where
\begin{enumerate}
    \item $Q$ is the finite set of states, containing  $\sum_i \text{dim}(S_i)$ elements, with dim($S_i$) equivalent  states corresponding to $S_i$ for each  $i\in\{1,\ldots,k\}$,
   \item $\Sigma$ is the same as the input alphabet of $M$,
    \item $q_1$ is the  initial state, selected arbitrarily from among the elements of $Q$ that correspond to the subspace containing the vector attained by $M$ after consuming the empty input string,
    \item   $F$ is the set of accepting states, designated to contain all and only the elements of $Q$ that correspond to any subspace containing  vectors attained by $M$ after consuming members of $L$, and
    \item  $\delta$ is the transition function, which mimics $M$'s action on its state vector, as follows: For each  $i\in\{1,\ldots,k\}$, call the subset of dim($S_i$) states corresponding to $S_i$ ``the $i$'th bag''. If $M$ maps vectors in $S_i$ to  vectors in $S_j$ upon reading a symbol $\sigma$, all states in the $i$'th bag of $M'$ will transition to states in the $j$'th bag with the symbol $\sigma$. For each bag, incoming transitions will be distributed as ``evenly'' as possible among the members of that bag, so that if $M'$ has a total of $T_j$ incoming $\sigma$-transitions to its $j$'th bag, no state in that bag  will have more than $\ceil{T_j/\text{dim}(S_j)}$ incoming $\sigma$-transitions.   
\end{enumerate}

Let us calculate the maximum possible number of incoming $\sigma$-transitions that can be received by a state in $M'$. Let $j$ be some state of $D$ with $p$ incoming $\sigma$-transitions from states  $\{i_1, i_2, ..., i_p\}$. For any $r\in\{1,2,...,p\}$, let $x$ be some string which carries $D$ to state $i_r$ and $M$ to some vector $\ket{\psi_x}\in \mathcal{H}_n$ with nonzero probability. We know that the processing of $\sigma$ corresponds to the action of the matrix we called $U_\sigma$ in Section \ref{sec:genQFA}. Recall from Equation \ref{eq:longvector} that
\[U_\sigma(\ket{\omega_1}\otimes \ket{\psi_{x}})= \ket{\omega_1}\otimes \widetilde{\ket{\psi_{x\sigma,1}}}+ \ket{\omega_2}\otimes \widetilde{\ket{\psi_{x\sigma,2}}} + \ldots + \ket{\omega_l}\otimes \widetilde{\ket{\psi_{x\sigma,l}}}.
\]
Since $M$ transitions to a vector in $S_j$ with probability 1 upon receiving $\sigma$, all the  $\widetilde{\ket{\psi_{x\sigma,k}}}$ must lie in $S_j$ (for $1\leq k\leq l$). We therefore have $U_\sigma(\ket{\omega_1}\otimes \ket{\psi_{x}})\subseteq 
\mathcal{H}_l\otimes S_j$. This is true for all $x$ and $\ket{\psi_x}$, and $S_{i_r}$ is, by definition, generated by such vectors; therefore, $U_\sigma(\mathbb{C}\ket{\omega_1}\otimes S_{i_r}) \subseteq 
\mathcal{H}_l\otimes S_j$.

By Corollary \ref{cor:subspaces}, the spaces $\mathbb{C}\ket{\omega_1}\otimes S_{i_r}$ are disjoint for all $r\in\{1,2,...,p\}$. We have
\begin{equation*}
    \text{dim}(\mathbb{C}\ket{\omega_1}\otimes S_{i_1}) + \text{dim}(\mathbb{C}\ket{\omega_1}\otimes S_{i_2}) + ... + \text{dim}(\mathbb{C}\ket{\omega_1}\otimes S_{i_p}) \leq \text{dim}(
    \mathcal{H}_l\otimes S_j),
\end{equation*}
since $U_\sigma$ is an injective linear map. 
In other words,
\begin{equation*}
    T_j = 
\text{dim}(S_{i_1}) + \text{dim}(S_{i_2}) + ... + \text{dim}(S_{i_p}) \leq l.\text{dim}(S_j).
\end{equation*}
Therefore, $\frac{T_j}{\text{dim}(S_j)} \leq l$, and no state receives more than $l$ incoming $\sigma$-transitions.
\end{proof}

We have seen that the erasure costs associated with the most expensive steps of zero-error QFAs are precisely representable by DFAs. This will let us conclude that the DFA-based results on these costs for the recognition of different regular languages to be established in the following section are valid for quantum machines as well.

\section{An upper bound for information erasure}\label{sec:hier}

Is there a universal bound on the number of bits that have to be ``forgotten'' by any computational step of any finite automaton? In this section, we provide an answer to this question.

\begin{theorem}\label{th:atmostnplus1}

Every regular language on an alphabet $\Sigma$  can be recognized by a DFA that has at most $\vert \Sigma \vert+1$ incoming transitions labeled with the same symbol to any of its states.

\end{theorem}
\begin{proof}
For unary input alphabets, any minimal machine is already in the required form.
For $k>1$, let $M$ be the minimal DFA recognizing some  language $L$ on the alphabet $\Sigma = \{\sigma_1,\dots, \sigma_k\}$. If $M$ is in the required form, we are done. Otherwise, let $Q=\{q_1, q_2, ..., q_n\}$ be the state set of $M$. 
We will add some new states (each of which  will be equivalent to some $q_i \in Q$) to $M$ to obtain a larger machine recognizing $L$. We will use the sets $C_1, C_2, ..., C_n$ so that each $C_i$ will contain the states that are equivalent to $q_i$ in our machine at every step of our construction.
Let $\mu_i$ denote the number of states in $C_i$ at any point in the process. Originally, each $C_i=\{q_i\}$, and each $\mu_i=1$.  Let $d_{\sigma,i}$ denote the total number of incoming transitions to states in $C_i$ with the symbol $\sigma$. Note that $d_{\sigma,i}$ is equal to the sum of $\mu_j$ over the $C_j$ containing states that transition to states in $C_i$ with the symbol $\sigma$, and it is independent of the details of which particular state in $C_i$ receives a particular $\sigma$-transition. Define
\begin{align}\label{definition_of_u_i}
u_i = \mu_i(k+1) - \max_{\sigma\in\Sigma}d_{\sigma,i}.
\end{align}
Note that if any state in any $C_i$ receives more than $k+1$ transitions with the same symbol, then the DFA does not satisfy the requirement of the theorem. The first term $\mu_i(k+1)$ is the maximum allowed number of incoming transitions with the same symbol to $C_i$ for the machine to  satisfy the requirement of the theorem. If each $u_i$ is nonnegative, then it is possible to distribute the incoming transitions among the states in $C_i$ such that no state receives more than $k+1$  transitions with the same symbol. The following algorithm finds $(\mu_i)_{i\in\{1, 2, ..., n\}}$ making all $u_i$ nonnegative.


\begin{itemize}
    \item Repeat until all $(u_i)_{1\leq i\leq n}$ are nonnegative:
    \begin{itemize}
        \item Find an index $i$ with the smallest $u_i$ value
        \item Add a state to $C_i$, increasing $\mu_i$ by 1
    \end{itemize}
\end{itemize}


At each iteration of the  loop, $\sum_{i\in\{1, ..., n\}}\mu_i(k+1)$ increases by $k+1$. On the other hand, $\sum_{i\in\{1, ..., n\}}\max_{\sigma\in\Sigma}d_{\sigma,i}$ also increases, but it increases by at most $k$, since the added state has only $k$ outgoing transition arrows. Therefore, by Equation \ref{definition_of_u_i}, the sum of all the $u_i$ increases  by at least one. Note that $\max_{i\in \{1, 2, ..., n\}}u_i$ is bounded above by a constant, because only those $u_i$ that are less than 0 are increased (by at most $k+1$) during the execution of the algorithm. The algorithm terminates at some point because the sum of finitely many bounded variables cannot increase by one forever.

The required DFA is then constructed by distributing each $C_i$'s incoming transitions evenly between its states, which works because each $u_i$ is now nonnegative.
\end{proof}

We now show that the bound shown in Theorem \ref{th:atmostnplus1}  is tight.
\begin{theorem}\label{th:atleastnplus1}

For every  $j\geq 1$, there exists a language $L_j$ on a $j$-symbol alphabet  with the following property: All DFAs recognizing $L_j$ have a state $q$ such that at least $j+1$ states transition to $q$ upon receiving the same symbol.

\end{theorem}
\begin{proof}
For the unary alphabet, it is easy to see that the language $L_1$ containing all strings except the empty string must have the property. For $j>1$, define the ``successor'' function $F$ on $\{1, ..., j\}$ by $F(i)=(i\text{ mod }j)+1$,
and let $B$ be $F$'s inverse. On the alphabet $\Sigma_j = \{\sigma_1, ..., \sigma_j\}$, define
\begin{align*}
    L_j = \{w |\; w \text{ ends with } \sigma_i\sigma_{F(i)} \text{ for some } 1\leq i\leq j\}.
\end{align*}

Let $M$ be a DFA recognizing $L_j$. Assume, without loss of generality, that $M$ does not have unreachable states.

Similarly to the proofs of Theorems \ref{th:qfatodfa} and 
\ref{th:atmostnplus1}, we will be using sets (``bags'') into which the states of $M$ are partitioned. Each bag contains states that are equivalent to the ones in the same bag, and distinguishable from all states in the other bags. $I$ is the bag that contains the initial state. For each $k$, $A_k$ is the bag containing the state reached by the input $\sigma_{B(k)}\sigma_k$, and $R_k$ is the bag containing any state reached by inputs of the form $\tau\sigma_k$, where $\tau$ is any substring not ending with $\sigma_{B(k)}$. 
Note that $A_i$ and $R_k$ are distinct bags for any $i,k\leq j$, because all states in $A_i$ are accepting states and those in $R_k$ are not. For $X \in \{A,R\}$, $X_k$ and $X_l$ are also distinct when $k\not=l$, since $M$ would reach an accepting state if it consumes the symbol  $\sigma_{F(k)}$ when in a member of $X_k$,  whereas it would reach a rejecting state with that symbol from a state in $X_l$. $I$ is distinct from all the $A_i$ and $R_i$, because it contains the only state which is two steps away from any accept state. The bags $(A_k)_k$, $(R_k)_k$ and $I$ partition the entire state set. 

The definition of $L_j$ dictates  that
all incoming transitions to states in $A_k$ or $R_k$ are labeled with the symbol $\sigma_k$. 
Let $i$ ($1\leq i\leq j$) be the index minimizing $|A_i|+|R_i|$, i.e. the sum of states in $A_i$ and $R_i$. Note that all states in all bags $(A_k)_k$, $(R_k)_k$ and $I$ transition to either $A_i$ or $R_i$ upon reading the symbol $\sigma_i$, so there are 
\begin{align*}
    \left(\sum_{1\leq k\leq j}|A_k|\right) + \left(\sum_{1\leq k \leq j}|R_k|\right) + |I|
\end{align*}
transitions with the symbol $\sigma_i$. Since $|A_i|+|R_i|$ is minimal and $|I|>0$, this number is strictly larger than $j(|A_i|+|R_i|)$. At least one state in $A_i$ or $R_i$ should thus have at least $j+1$ incoming $\sigma_i$-transitions by the pigeonhole principle.
\end{proof}

 Theorems \ref{th:atmostnplus1}
and \ref{th:atleastnplus1} imply that, for any particular temperature $T$, given any amount of energy, there exists a regular language (on a suitably large alphabet) whose 
recognition at $T$ requires a DFA with at least that much energy cost for at least one of its computational steps. When the alphabet is fixed, one can always rewrite any DFA on that alphabet to obtain a  machine recognizing the same language with each step costing no more than the bound proven in Theorem \ref{th:atmostnplus1}. By Theorem \ref{th:qfatodfa}, the same energy costs are associated with zero-error QFAs for that language.

\section{Trading energy for error }\label{sec:boundederror}

It turns out that the minimum energy required for the most expensive step during the recognition of some regular languages is reduced if   one allows the finite automaton to give erroneous answers with probability not exceeding some bound less than  $\frac{1}{2}$. 

Recall the language family $\{L_j 
| j\geq 1\}$ defined in the proof of Theorem \ref{th:atleastnplus1}. Any zero-error QFA recognizing  some $L_j$ must have at least $j+1$ operating elements by Theorem \ref{th:qfatodfa}. Since $L_1$ is not a group language, no QFA with a single operating element can recognize it, even with bounded error \cite{BP02}. 

\begin{theorem}\label{th:just2}
 There exists a QFA  with  two operating elements per superoperator that recognizes the language $L_2$ with bounded error.
\end{theorem}
\begin{proof}
Consider Fig. \ref{fig:QFAdiag}, which depicts the transitions of a QFA named $M_2$. (All arrows in the figure correspond to transitions with amplitude 1, unless otherwise indicated.) The superoperators for the left end-marker and the two input symbols are shown in Fig. \ref{fig:QFAmatrix2}.  Upon reading the 
left end-marker,  $M_2$ branches to three equal-amplitude submachines that never interfere in the remainder of the computation. For $i \in \{1,2\}$, submachine $M_{2,i}$ accepts an input string if and only if it  ends with  $\sigma_i\sigma_{F(i)}$.   Submachine $M_{2,3}$ accepts every input. Since any string is in $L_2$ if and only if it is accepted by one of $M_{2,1}$ and $M_{2,2}$, $M_2$ recognizes $L_2$ with error probability $\frac{1}{3}$.
\end{proof}


\begin{theorem}\label{th:error}
For all $j \geq 3$, there exists a QFA $M_j$ with  three operating elements per superoperator that recognizes the language $L_j$ with error probability bounded by $\frac{j-1}{2j-1}$. 
\end{theorem}

\begin{figure}
    \centering
\scalebox{0.55}{
        \begin{tikzpicture}
        
             \node[state, initial] (one) {1};
             \node[state, above=1.2 of one, xshift=2cm] (two) {2};
             \node[state, above=1.2 of one, xshift=4cm] (three) {3};
             \node[state, above=1.2 of one, accepting, xshift=6cm] (four) {4};
             \node[state, above=1.2 of one, xshift=8cm] (five) {5};
             \node[state, xshift=2cm] (six) {6};
             \node[state, xshift=4cm] (seven) {7};
             \node[state, accepting, xshift=6cm] (eight) {8};
             \node[state, xshift=8cm] (nine) {9};
             \node[state, below=1.2 of one, accepting, xshift=2cm] (ten) {10};
             
             \node at (9.6,1.75) {\huge{\} $M_{2,1}$}};
             \node at (9.6,0) {\huge{\} $M_{2,2}$}};
             \node at (4.6,-1.9) {\huge{\} $M_{2,3}$}};
             
            \draw   (one) edge[bend left, above, ->] node[rotate=40] {\texttt{$\lend, \frac{1}{\sqrt{3}}$}} (two)
                    (one) edge[above, ->] node{\texttt{$\lend, \frac{1}{\sqrt{3}}$}} (six)
                    (one) edge[bend right, below, ->] node[rotate=-40] {\texttt{$\lend, \frac{1}{\sqrt{3}}$}} (ten)

                    (two) edge[below, ->] node{\texttt{$\sigma_1$}} (three)
                    (three) edge[below, ->] node{\texttt{$\sigma_2$}} (four)
                    (four) edge[below, ->] node{\texttt{$\sigma_1$}} (five)
                    (two) edge[loop above] node{\texttt{$\sigma_2$}} (two)
                    (three) edge[loop above, ->] node{\texttt{$\sigma_1$}} (three)
                    (four) edge[bend right=65, above, ->] node{\texttt{$\sigma_2$}} (two)
                    (five) edge[bend right, above, ->] node{\texttt{$\sigma_2$}} (four)
                    (five) edge[loop above] node{\texttt{$\sigma_1$}} (five)

                    (six) edge[below, ->] node{\texttt{$\sigma_2$}} (seven)
                    (seven) edge[below, ->] node{\texttt{$\sigma_1$}} (eight)
                    (eight) edge[below, ->] node{\texttt{$\sigma_2$}} (nine)
                    (six) edge[loop above] node{\texttt{$\sigma_1$}} (six)
                    (seven) edge[loop above, ->] node{\texttt{$\sigma_2$}} (seven)
                    (eight) edge[bend left=40, above, ->] node{\texttt{$\sigma_1$}} (six)
                    (nine) edge[bend right, above, ->] node{\texttt{$\sigma_1$}} (eight)
                    (nine) edge[loop above] node{\texttt{$\sigma_2$}} (nine)
                    
                    (ten) edge[loop right] node{\texttt{$\sigma_1$}} (ten)
                    (ten) edge[loop below] node{\texttt{$\sigma_2$}} (ten);
        \end{tikzpicture}
    }    
    \caption{A QFA recognizing $L_2$ with bounded error} 
    \label{fig:QFAdiag}
\end{figure}

\begin{figure}
     \begin{subfigure}{.3\textwidth}
     \centering
        \includegraphics[width=1\textwidth]{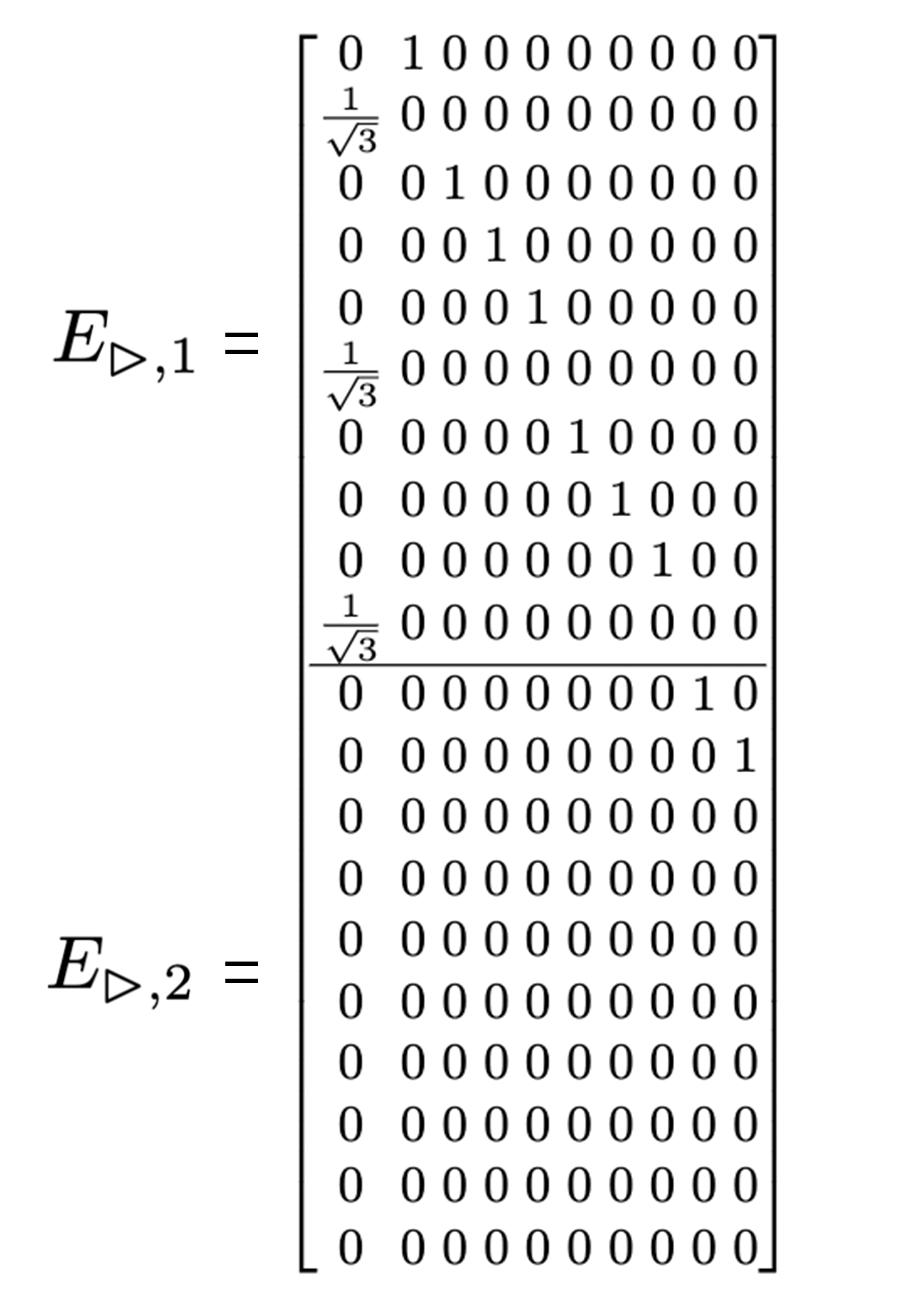}
        \caption{Superoperator for \lend}
     \end{subfigure}
     \begin{subfigure}{.3\textwidth}
     \centering
        \includegraphics[width=1\textwidth]{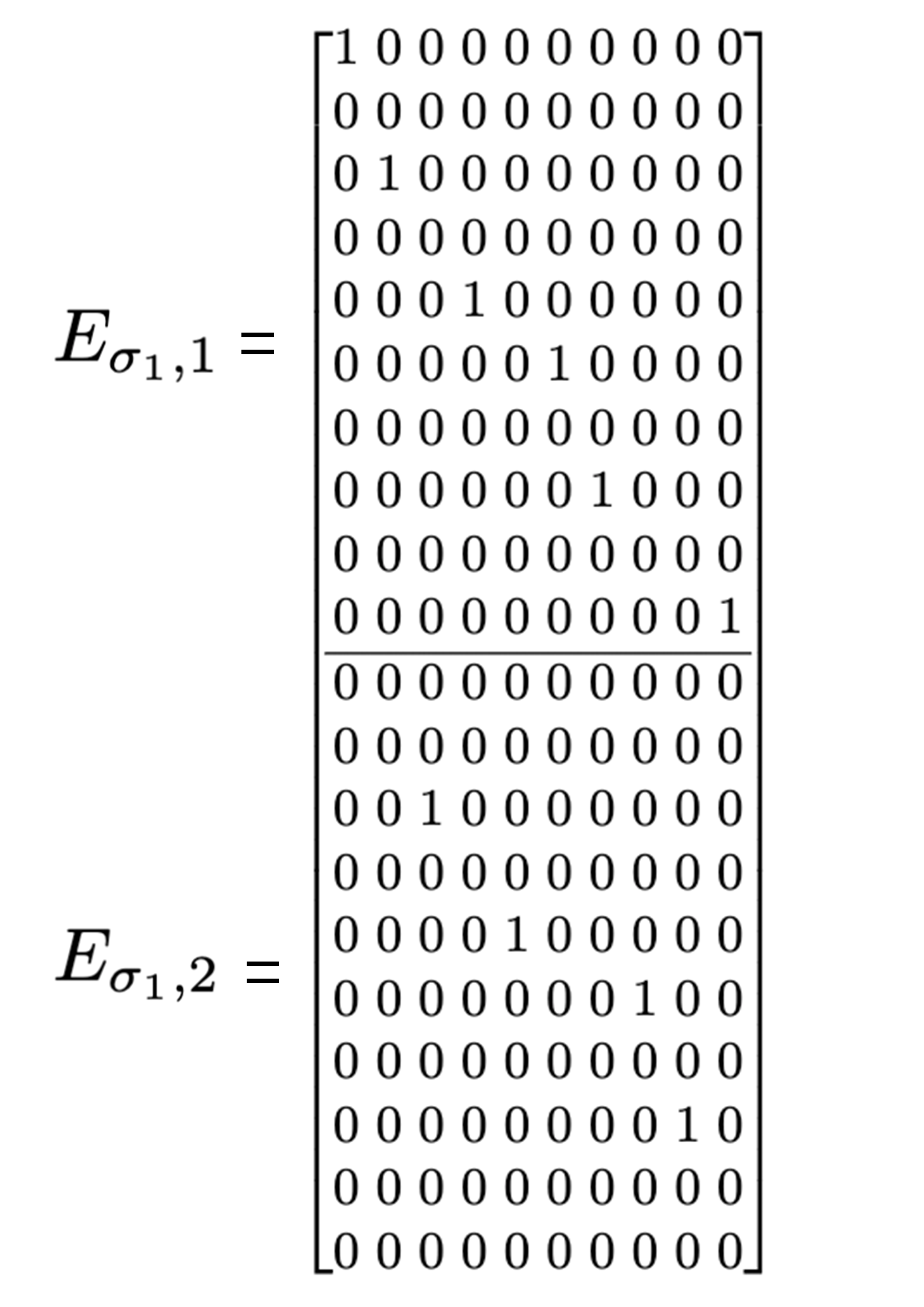}
        \caption{Superoperator for $\sigma_1$}
     \end{subfigure}
     \begin{subfigure}{.3\textwidth}
     \centering
        \includegraphics[width=1\textwidth]{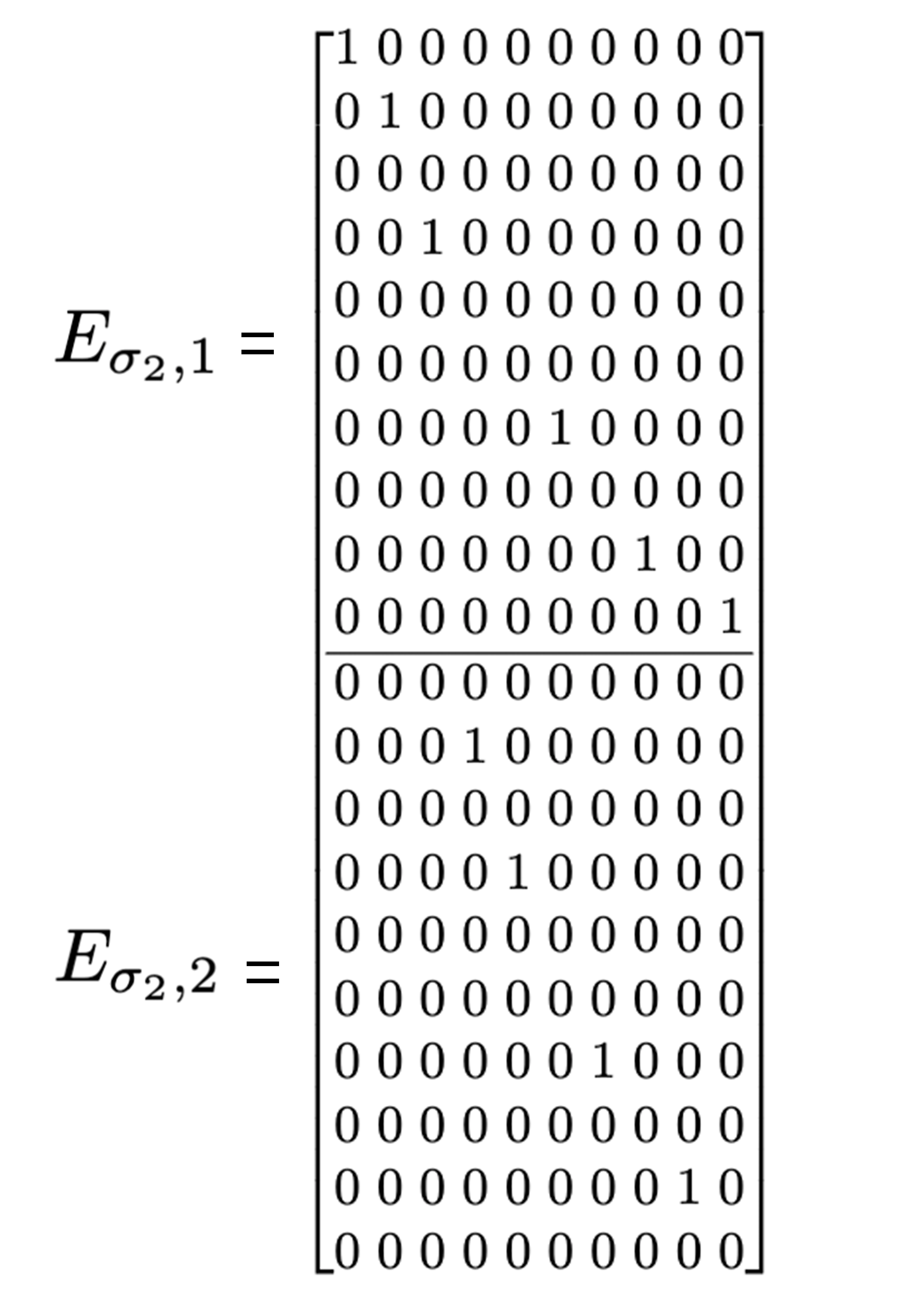}
        \caption{Superoperator for $\sigma_2$}
     \end{subfigure}
     \caption{Transition matrices for the QFA of Figure \ref{fig:QFAdiag}}
     \label{fig:QFAmatrix2}
 \end{figure}

\begin{figure}
    \centering
    \scalebox{0.85}{
        \begin{tikzpicture}[initial text={$\lend, \frac{1}{\sqrt{2j-1}}$}]
        
             \node[state] (one) {};
             \node[state, xshift=2.5cm] (two) {};
             \node[state, accepting, xshift=4.5cm] (three) {};
             
            \draw   (one) edge[below, ->] node{\texttt{$\sigma_i$}} (two)
                    (two) edge[above, ->] node{\texttt{$\sigma_{F(i)}$}} (three)
                    (three) edge[bend left=50, below, ->] node{\texttt{$\sigma_i$}} (two)
                    (two) edge[bend right, above, ->] node{\texttt{$\Sigma_j-\{ \sigma_i,\sigma_{F(i)}\}$}} (one)
                    (three) edge[bend right=100, above, ->] node{\texttt{$\Sigma_j-\{\sigma_i\}$}} (one)
                    (one) edge[loop below, ->] node{\texttt{$\Sigma_j-\{\sigma_i\}$}} (one)
                    (two) edge[loop below, ->] node{\texttt{$\sigma_i$}} (two)
                    (-2,0) edge[above, ->] node{\texttt{$\lend, \frac{1}{\sqrt{2j-1}}$}} (one);
        \end{tikzpicture}
        }
    \caption{Submachine $M_{j,i}$ in the construction of Theorem \ref{th:error}} 
    \label{fig:subdiag}
\end{figure}

\begin{proof}
The argument is similar to the one in the proof of Theorem \ref{th:just2}. Machine $M_j$ has $j+1$ submachines.
For $i \in \{1,\dots,j\}$, submachine $M_{j,i}$ (depicted in Fig. \ref{fig:subdiag}) accepts its input  if and only if it  ends with  $\sigma_i\sigma_{F(i)}$, whereas   submachine $M_{j,j+1}$ accepts every input. (In Fig. \ref{fig:subdiag}, arrow labels of the form $\Sigma_j - \Gamma$ express that all symbols of the input alphabet except those in set $\Gamma$ effect a transition with amplitude 1 between the indicated states.)
$M_j$ starts by branching with amplitude $\frac{1}{\sqrt{2j-1}}$ to each of $M_{j,i}$ for  $i \in \{1,\dots,j\}$, and with amplitude $\frac{\sqrt{j-1}}{\sqrt{2j-1}}$ to $M_{j,j+1}$.
Strings in $L_j$ must lead  one of the first $j$ submachines to acceptance, and ``tip the balance'' for the overall machine to accept with probability $\frac{j}{2j-1}$. It is easy to see in Fig. \ref{fig:subdiag} that the superoperators would have just three operating elements.
\end{proof}




\section{Energy complexity of languages}\label{sec:langcost}

In complexity theory, it is standard to represent the worst-case amount of a computational resource used by an algorithm as a function of the length of its input. A lower bound on the growth rate of this function for any algorithm for a specific problem then serves as a measure of the ``difficulty'' of that problem. Let us examine how our new  energy cost metric can be used to classify regular languages in this fashion. Recall that we define the energy complexity of a finite automaton as the maximum number of bits that it can delete when running on any input string of  length $n$.

Group languages \cite{BP02} have DFAs that contain no states with more than one incoming transition with the same input symbol, and, equivalently, QFAs that have a single operation element for each symbol. These languages can be recognized with zero energy cost. We now proceed to analyze the remaining case, that is, the energy requirements of  irreversible regular languages.

\begin{theorem}\label{th:linenerg}
Every irreversible regular language has linear energy complexity.
\end{theorem}

The proof of Theorem \ref{th:linenerg} consists of the following two lemmas, which establish the lower and upper bounds, respectively.

\begin{lemma}\label{lemma:lowerb}
Every DFA recognizing an irreversible language has energy complexity $\Omega(n)$.
\end{lemma}
\begin{proof}
Let $M$ be a DFA that recognizes an irreversible language $L$. Without loss of generality, assume that all states of $M$ are reachable. We claim that the transition function of $M$ contains a ``loop'' that allows some state with multiple incoming transitions with the same symbol to be visited arbitrarily many times for sufficiently long inputs. 

Since $L$ is not reversible, there exists at least one state, say, $q$, of $M$ which has at least two incoming transitions labeled with some input symbol $\sigma$.
Assume that the input string $w$ brings $M$ to $q$ from the start state.

Note that a sufficiently long input string of the form $w\sigma^*$ must eventually ``trap'' $M$ in a loop of $U\geq 1$ states linked with $\sigma$-transitions. If $q$ is a member of that loop, we have proven our claim. Otherwise, some other state $q'$ that is reachable from $q$ by a substring of one or more $\sigma$'s must be the ``entry point'' to the loop, and must receive at least two $\sigma$-transitions that originate from one state in the loop and one state outside the loop. This proves the claim.

Consider the energy expenditure of $M$ on inputs of the form $w\sigma^*$. It is clear that the repeated state will have to be visited once every $U$ steps for long input strings, leading to an energy complexity  that is proportional to at least $\frac{n}{U}$.
\end{proof}

As for the upper bound, 
it is evident that each finite automaton $M$ has energy complexity $c_M n$ for some constant $c_M$: For any input alphabet $\Sigma$, the bound proven in Theorem \ref{th:atmostnplus1} shows that no more than $\log_2(|\Sigma|+1)n$ bits need to be erased by a machine with that alphabet on any input of length $n$. The following lemma brings this bound lower.

\begin{lemma}\label{lemma:upperb}
For every regular language $L$ on alphabet $\Sigma$ and every positive real number $\epsilon$, there exists a finite automaton with energy complexity at most $(\log_2|\Sigma|+\epsilon)n$ that recognizes $L$.
\end{lemma}
\begin{proof}
The idea is to trade memory (i.e. number of states) for energy: Let $M_0$ be the minimal-state DFA recognizing $L$ with $C$ states. We consider equivalent machines with more and more states: For any integer $k > 0$, consider a  machine $M_k$ with 
which imitates $M_0$ while scanning successive $(k+1)$-symbol substrings of the input and then ``forgetting'' the information related to them wholesale once every $k+1$ steps of the computation, rather than at every step. Fig. \ref{fig:trees} illustrates this construction for the language recognized by the DFA of Fig. \ref{fig:sub1} for $k=3$. The root of each tree in the diagram in that figure corresponds to a point where the machine starts processing a new segment of the input with its memory containing nothing about the consumed input prefix $w$ except the  state that $M_0$ would be at after consuming $w$. Each state is labeled with the name of the state that it mimics in the diagram of Fig. \ref{fig:sub1}, and each ``missing'' transition from any state at the lowest level of a tree enters the root of the tree corresponding to the target of the imitated transition in Fig. \ref{fig:sub1}.

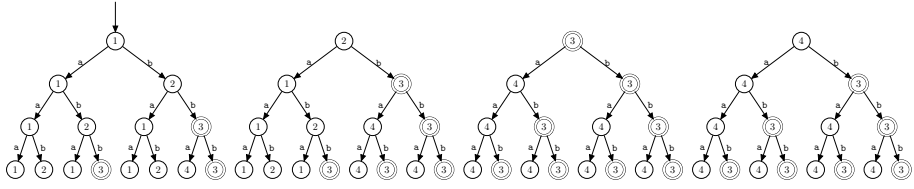
\begin{figure}
\centering
\scalebox{0.38}{
\begin{tikzpicture}[level distance=1.5cm,  level 1/.style={sibling distance=4cm},  level 2/.style={sibling distance=2cm},  level 3/.style={sibling distance=1cm},  edge from parent/.style={->,draw}, acceptingstate/.style={double distance=2pt}]

\begin{scope}[level 2/.style={sibling distance=4cm},  level 3/.style={sibling distance=2cm},  level 4/.style={sibling distance=1cm}]
    \node {} 
        child {
            node[circle,draw] {1}
              child {
                node[circle,draw] {1} 
                child {
                  node[circle,draw] {1}
                  child {node[circle,draw] {1}edge from parent node[left] {\texttt{a}}}
                  child {node[circle,draw] {2}edge from parent node[right] {\texttt{b}}}
                edge from parent node[left] {\texttt{a}}}
                child {
                  node[circle,draw] {2}
                  child {node[circle,draw] {1}edge from parent node[left] {\texttt{a}}}
                  child {node[circle,draw,acceptingstate] {3}edge from parent node[right] {\texttt{b}}}
                edge from parent node[right] {\texttt{b}}}
              edge from parent node[left] {\texttt{a}}}
              child {
                node[circle,draw] {2}
                child {
                  node[circle,draw] {1}
                  child {node[circle,draw] {1} edge from parent node[left] {\texttt{a}}}
                  child {node[circle,draw] {2} edge from parent node[right] {\texttt{b}}}
                edge from parent node[left] {\texttt{a}}}
                child {
                  node[circle,draw,acceptingstate] {3}
                  child {node[circle,draw] {4}edge from parent node[left] {\texttt{a}}}
                  child {node[circle,draw,acceptingstate] {3}edge from parent node[right] {\texttt{b}}}
                edge from parent node[right] {\texttt{b}}}
            edge from parent node[right] {\texttt{b}}}
        };
\end{scope}

\begin{scope}[xshift=8cm, yshift=-1.5cm]
\node[circle,draw] {2}
  child {
    node[circle,draw] {1}
    child {
      node[circle,draw] {1}
      child {node[circle,draw] {1}edge from parent node[left] {\texttt{a}}}
      child {node[circle,draw] {2}edge from parent node[right] {\texttt{b}}}
    edge from parent node[left] {\texttt{a}}}
    child {
      node[circle,draw] {2}
      child {node[circle,draw] {1}edge from parent node[left] {\texttt{a}}}
      child {node[circle,draw,acceptingstate] {3}edge from parent node[right] {\texttt{b}}}
    edge from parent node[right] {\texttt{b}}}
  edge from parent node[left] {\texttt{a}}}
  child {
    node[circle,draw,acceptingstate] {3}
    child {
      node[circle,draw] {4}
      child {node[circle,draw] {4}edge from parent node[left] {\texttt{a}}}
      child {node[circle,draw,acceptingstate] {3}edge from parent node[right] {\texttt{b}}}
    edge from parent node[left] {\texttt{a}}}
    child {
      node[circle,draw,acceptingstate] {3}
      child {node[circle,draw] {4}edge from parent node[left] {\texttt{a}}}
      child {node[circle,draw,acceptingstate] {3}edge from parent node[right] {\texttt{b}}}
    edge from parent node[right] {\texttt{b}}}
  edge from parent node[right] {\texttt{b}}};
\end{scope}
\begin{scope}[xshift=16cm, yshift=-1.5cm]
\node[circle,draw,acceptingstate] {3}
  child {
    node[circle,draw] {4}
    child {
      node[circle,draw] {4}
      child {node[circle,draw] {4}edge from parent node[left] {\texttt{a}}}
      child {node[circle,draw,acceptingstate] {3}edge from parent node[right] {\texttt{b}}}
    edge from parent node[left] {\texttt{a}}}
    child {
      node[circle,draw,acceptingstate] {3}
      child {node[circle,draw] {4}edge from parent node[left] {\texttt{a}}}
      child {node[circle,draw,acceptingstate] {3}edge from parent node[right] {\texttt{b}}}
    edge from parent node[right] {\texttt{b}}}
  edge from parent node[left] {\texttt{a}}}
  child {
    node[circle,draw,acceptingstate] {3}
    child {
      node[circle,draw] {4}
      child {node[circle,draw] {4}edge from parent node[left] {\texttt{a}}}
      child {node[circle,draw,acceptingstate] {3}edge from parent node[right] {\texttt{b}}}
    edge from parent node[left] {\texttt{a}}}
    child {
      node[circle,draw,acceptingstate] {3}
      child {node[circle,draw] {4}edge from parent node[left] {\texttt{a}}}
      child {node[circle,draw,acceptingstate] {3}edge from parent node[right] {\texttt{b}}}
    edge from parent node[right] {\texttt{b}}}
  edge from parent node[right] {\texttt{b}}};
\end{scope}
\begin{scope}[xshift=24cm, yshift=-1.5cm]
\node[circle,draw] {4}
  child {
    node[circle,draw] {4}
    child {
      node[circle,draw] {4}
      child {node[circle,draw] {4}edge from parent node[left] {\texttt{a}}}
      child {node[circle,draw,acceptingstate] {3}edge from parent node[right] {\texttt{b}}}
    edge from parent node[left] {\texttt{a}}}
    child {
      node[circle,draw,acceptingstate] {3}
      child {node[circle,draw] {4}edge from parent node[left] {\texttt{a}}}
      child {node[circle,draw,acceptingstate] {3}edge from parent node[right] {\texttt{b}}}
    edge from parent node[right] {\texttt{b}}}
  edge from parent node[left] {\texttt{a}}}
  child {
    node[circle,draw,acceptingstate] {3}
    child {
      node[circle,draw] {4}
      child {node[circle,draw] {4}edge from parent node[left] {\texttt{a}}}
      child {node[circle,draw,acceptingstate] {3}edge from parent node[right] {\texttt{b}}}
    edge from parent node[left] {\texttt{a}}}
    child {
      node[circle,draw,acceptingstate] {3}
      child {node[circle,draw] {4}edge from parent node[left] {\texttt{a}}}
      child {node[circle,draw,acceptingstate] {3}edge from parent node[right] {\texttt{b}}}
    edge from parent node[right] {\texttt{b}}}
  edge from parent node[right] {\texttt{b}}};
\end{scope}

\end{tikzpicture}
}
\caption{An ``energy-efficient'' DFA equivalent to the machine of Fig. \ref{fig:sub1} (See text for the unspecified transitions.)}
\label{fig:trees}
\end{figure}

For inputs of length $n$, the automaton $M_k$ executes at most $\lfloor\frac{n}{k+1}\rfloor$ steps that can have positive energy cost. The cost of any such step is bounded above by the logarithm of the maximum possible number  of incoming $\sigma$-transitions to a root state for any symbol $\sigma$, which is just $\log_2(C|\Sigma|^k)$. The energy complexity of $M_k$ is therefore at most $\log_2(C|\Sigma|^k)\lfloor n/(k+1)\rfloor$, that is,
\[
\log_2(C)\lfloor n/(k+1)\rfloor+\log_2(|\Sigma|^k)\lfloor n/(k+1)\rfloor<
\left(\log_2|\Sigma|+\frac{\log_2(C)}{k+1}\right)n,
\]
where one can  make the second term in the parentheses as small as one likes by choosing $k$ accordingly.
\end{proof}
 
\begin{figure}[htp]
    \centering
    \scalebox{0.85}{
        \begin{tikzpicture}[initial text={}]
        
             \node[state] (one) {};
             \node[state, accepting, xshift=2.2 cm] (two) {};
             
            \draw   (one) edge[below, ->] node{\texttt{$\Sigma$}} (two)
                        (two) edge[loop above, ->] node{\texttt{$\Sigma$}} (two)
                        (-1,0) edge[above, ->] node{\texttt{}} (one);
        \end{tikzpicture}
        }
    \caption{The minimal DFA recognizing $L_I$} 
    \label{fig:LI}
\end{figure}
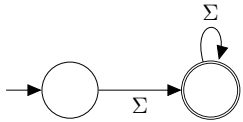

 We aim to discover a finer classification of regular languages   by examining, for each language $L$, a lower bound on the constant $c_M$ on any automaton $M$ that recognizes $L$ with energy complexity $c_M n$. 
 In this context, consider the following nonempty subclass: Define an irreversible language $L$ to be \textit{minimally expensive} if, for  every positive real number $\epsilon$, $L$ can be recognized by some finite automaton $M$ with energy complexity $c_M n$, where $c_M < \epsilon$. As an example, consider the language $L_I=\{w | w\in\Sigma^*, |w|\geq 1\}$. The minimal DFA recognizing $L_I$ (Fig. \ref{fig:LI}), erases one bit of information at each step of its execution, as it repeatedly ``forgets''  which state it left in the previous transition.

\begin{figure}[htp]
    \centering
    \scalebox{0.85}{
        \begin{tikzpicture}[initial text={}]
        
             \node[state] (one) {};
             \node[state, accepting, xshift=2 cm] (two) {};
             \node[state, accepting, xshift=4 cm] (three) {};
             \node[state, accepting, xshift=6 cm] (four) {};
            \draw   (one) edge[below, ->] node{\texttt{$\Sigma$}} (two)
                        (two) edge[below, ->] node{\texttt{$\Sigma$}} (three)
                        (three) edge[below, ->] node{\texttt{$\Sigma$}} (four)
                        (four) edge[bend right=45, above, ->] node{\texttt{$\Sigma$}} (two)
                        (-1,0) edge[above, ->] node{\texttt{}} (one);
        \end{tikzpicture}
        }
    \caption{A more energy-efficient DFA for $L_I$} 
    \label{fig:betterLI}
\end{figure}
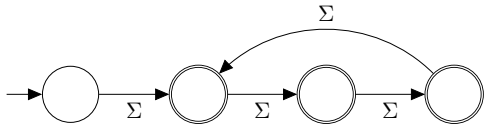

The machine of Fig. \ref{fig:LI} is, however, far from being the most ``energy-efficient'' finite automaton for $L_I$. Consider what happens when one replaces the looping state in Fig. \ref{fig:LI} with a greater collection of equivalent states, as in Fig. \ref{fig:betterLI}: Only one of the states in the new, bigger loop involves forgetting where one came from, and paying the energy bill necessitated with that erasure. Like the construction in the proof of Lemma \ref{lemma:upperb}, this new machine with three states in the loop uses additional memory to store the information to be erased  for a longer time, and spends only a third of the energy spent by the previous machine for identical input strings. Since one can decrease the coefficient of $n$ in the energy complexity associated with the recognition of $L_I$ to any desired positive value by simply increasing  the number of states in the loop, $L_I$ is a minimally expensive language.


We now show that some irreversible languages are not minimally expensive, by demonstrating that any DFA recognizing them must in fact forget more information than contained in the input string for certain long inputs, i.e. that the bound of Lemma \ref{lemma:upperb} is tight for those languages. In the following, let $\chi(q, \sigma)$ denote the number of incoming $\sigma$-transitions to state $q$. We are  interested in the quantity
$\sum_{i=1}^{n}\log_2(\chi(u_i, \sigma_i))$, which represents the total information forgotten when our DFA consumes an $n$-bit string $\sigma_1 \sigma_2...\sigma_n$ and traverses the state sequence $u_1, u_2, ..., u_n$, starting from some state $u_0$. Let $L = \{w\text{ } |\text{ } w \text{ contains the substring } \texttt{bb} \text{ and ends with a } \texttt{b}\}$, the minimal DFA for which is shown in Fig. \ref{fig:sub1}.


\begin{theorem}
Let $M$ be any DFA recognizing the language $L$. Then there exists $\epsilon>0$ such that for all sufficiently large $n$, there exists an input string $w=w_1 w_2...w_n$ for which
\[
    \sum_{i=1}^{n}\log_2(\chi(q_i, w_i)) \geq (1+\epsilon)n,
\]
where  $(q_0, q_1, q_2, ..., q_n)$ is the sequence of states traversed by $M$ during the consumption of $w$, beginning with the start state $q_0$.
\end{theorem}

\begin{proof}
Assume without loss of generality that $M$ has no unreachable states. Note that no state of $M$ can have different incoming transitions associated with different input symbols. For every state $u$ of $M$ with at least one incoming transition with some symbol $\sigma$, assign unique labels from the set $T=\{1, 2, ... , \chi(u, \sigma)\}$ to those transitions in some arbitrary fashion.

Define an \textit{$n$-step subcomputation} of $M$ to be a sequence of moves (not necessarily starting from the start state) during which it consumes a sequence of $n$ symbols. Each such subcomputation is associated with the sequence of traversed states $\mathbf{u}=(u_0, u_1, u_2, ..., u_n)$, the sequence of consumed symbols $\mathbf{s}=(\sigma_1,\sigma_2, ..., \sigma_n)$, and the sequence of the labels of the used transitions, say,  $\mathbf{t}=(t_1, t_2, ..., t_n)$.
Note that a particular $n$-step subcomputation is uniquely identified by the pair $(u_0,\mathbf{s})$;  all the other items mentioned above can be determined by ``simulating'' $M$ if one knows this pair. Similarly, the pair $(u_n,\mathbf{t})$ also determines everything about an $n$-step subcomputation, since one can ``go'' from $u_n$ all the way back to $u_0$ via transitions labeled $t_n, t_{n-1}$, etc. without any ambiguity about the symbol consumed at any step, due to the property of $M$ we noted at the beginning.

Consider a scenario where we feed $M$ input symbols selected uniformly randomly from $\{\texttt{a},\texttt{b}\}$ at each successive step. Under these conditions, a DFA will behave like a Markov chain, and is guaranteed to have a stationary distribution. \cite{H10}  Let $D$ be such a distribution for $M$. It is clear from Fig. \ref{fig:sub1} that  each state of $M$ with nonzero probability in $D$ is equivalent to some state in $\{3,4\}$, with each of those classes having at least one positive-probability state in $D$.  


Consider an information source which repeatedly transmits  $n$-step subcomputations of $M$ that it chooses by picking $u_0$ from $D$ and $\mathbf{s}$ uniformly randomly. The Shannon entropy of this source is $n+H(u_0)$, where $H(u_0)$ is the  entropy of $D$, which is a lower bound for the number of bits required for expressing $u_0$. Note that the entropies associated with all the states $u_1, ..., u_n$ in this subcomputation  also equal $H(u_0)$, since $D$ is stationary.

Let us design a schema for coding the messages of this information source. One idea is to send a pair $(u_n,\mathbf{t})$, which, as mentioned above, uniquely identifies the subcomputation. One could start the message by describing $u_n$, and continue with a concatenation of the binary encodings of the labels $t_n, t_{n-1}$, etc. 
However, it is well known that the average number of bits per message is reduced if one ``compresses''  several such messages together into a single ``package''. Consider the following, more sophisticated scheme for packing some large number $p$ of messages together in such a fashion.

The bit string we construct begins describing $p$ pairs of the form $(u_n,\mathbf{t})$ with a prefix that describes  the first components of each pair by listing $p$ state names. 
We use  Huffman coding (although any other similarly efficient code would also do) and our knowledge of $D$ to encode this information. Let us call the length of this prefix $U$.

It remains to encode the label names $\runo {t^{1}} n, \runo {t^2} n, ..., \runo {t^p} n$ appearing in the $p$ sequences of transition labels corresponding to the $p$ subcomputations we wish to describe. We will group these labels according to the in-degrees of the states into which they enter in the transition diagram of $M$. States with in-degree 1 will be ignored, since one needs no information to ``rewind'' one step from such a state. For each   $i>1$, let $S_i^j$ be the number of transitions into states with in-degree $i$ in the $j$th subcomputation in the package. Let $S_i = \sum_{j=1}^p S_i^j$. Consider ``filtering''  the sequence $\runo {t^{1}} n, \runo {t^2} n, ..., \runo {t^p} n$ so that one is left with just the 
 $S_i$ labels of the transitions entering states with this in-degree. Note that this possibly shorter sequence can be viewed as an $S_i$-digit number $N_i$ written in base $i$. Each such $N_i$ will be specified in a dedicated block, as described below.

Let $\tau$ be the maximum in-degree in the state diagram of $M$. For each $i$ from 2 up to $\tau$, the description of the labels of transitions into states with in-degree $i$ begins with a     $\ceil{\log_2(pn)}$-bit prefix encoding the number $S_i$, and ends with a $\ceil{\log_2 i^{S_i}}$-bit postfix, which is the binary encoding of $N_i$. We concatenate these $\tau-1$ descriptions to the description of the $p$ final states mentioned earlier. This concludes our bit string encoding $p$ $n$-step subcomputations.  

The length of this package is
    $(\tau-1)\ceil{\log_2(pn)} + \sum_{i=2}^{\tau}\ceil{\log_2(i^{S_i})} + U$.
This  is a random variable, since the  second and third terms depend on the randomly selected subcomputations in the package.
The expected length of the code for each subcomputation is

\begin{align*}
     &\frac{(\tau-1)\ceil{\log_2(pn)}}{p} + \sum_{i=2}^{\tau}\frac{\mathbb{E}[\ceil{\log_2(i^{S_i})}]}{p} +\mathbb{E}\left[\frac{U}{p}\right],
\end{align*}
which is at most
\begin{equation}\label{eq:upb1}
    \frac{(\tau-1)\log_2(pn)}{p} + \frac{\sum_{i=2}^{\tau}\log_2(i)\mathbb{E}[S_i]}{p} + \frac{\mathbb{E}[U]}{p} + \frac{2\tau}{p}.
\end{equation}
Note that $S_i$ is the sum of $S_i^j$, all of which have the same distribution (and the same expectation). We have
\begin{align*}
    &\frac{\sum_{i=2}^{\tau}\log_2(i)\mathbb{E}[S_i]}{p} = \frac{\sum_{i=2}^{\tau}\log_2(i)\sum_{j=1}^p\mathbb{E}[S_i^j]}{p}\\ = &\frac{\sum_{i=2}^{\tau}\log_2(i)\sum_{j=1}^p\mathbb{E}[S_i^1]}{p} = \frac{\sum_{i=2}^{\tau}\log_2(i)p\mathbb{E}[S_i^1]}{p}\\
    = &\sum_{i=2}^{\tau}\log_2(i)\mathbb{E}[S_i^1] = \sum_{i=2}^{\tau}\mathbb{E}[S_i^1\log_2(i)] \\
    = &\mathbb{E}\left[\sum_{i=2}^{\tau}S_i^1\log_2(i)\right].
\end{align*}

It is important to see that this last value equals $\mathbb{E}[\codinglength]$: These expressions are two different ways of denoting the expectation of a random variable which is the sum of the logarithms of the in-degrees of the $n$ states traversed in a randomly selected $n$-step subcomputation of $M$. One sum lists these terms in ``chronological'' order, whereas the other one groups them according to the associated in-degree. 

We can thus rewrite the upper bound (\ref{eq:upb1}) we obtained above for the expected number of bits per subcomputation as
\begin{equation}\label{eq:upb2}
    \mathbb{E}\left[\codinglength\right] +
    \frac{\mathbb{E}[U]}{p} +
    \frac{(\tau-1)\log(pn)}{p} +
    \frac{2\tau}{p}.
\end{equation}
In expression (\ref{eq:upb2}),  the last two terms tend to $0$ as $p$ grows. The second term approaches $H(u_0)$, which is the Shannon entropy of the probability distribution of a ``final state'' in a randomly selected subcomputation, as discussed above. 
Our code is self-delimiting and uniquely decodable. By Shannon's source coding theorem,  the expected length of our code for a subcomputation cannot be less than the  entropy of the source, which we know to be $n+H(u_0)$, so
\begin{equation*}
    n+H(u_0)
    \leq 
 \mathbb{E}\left[\codinglength\right]
    + H(u_0),
\end{equation*}
which yields
\begin{equation}\label{eq:chili}
    \mathbb{E}\left[\codinglength\right]
    \geq n.
\end{equation}

We now propose an even more efficient way to encode our subcomputations. Note that no subcomputation selected by our source will ever traverse a transition emanating from a state which has zero probability in $D$. Knowing $D$, one can ignore all those transitions and still specify a path in the state diagram that visits only the positive-probability states. For every such state $q$, let $\psi(q, \sigma)$ denote the number of  $\sigma$-transitions incoming to state $q$ from states with positive probability in $M$. We note that $1\leq\psi(q, \sigma)\leq\chi(q, \sigma)$ for all such $q$. Furthermore, there exist at least one state $q'$ and a symbol $\gamma$ for which $\psi(q', \gamma)<\chi(q', \gamma)$, since the initial state of $M$ has zero probability in $D$ and so there must be at least one transition which allows $M$ to move from a zero-probability state to one with positive probability. Our latest code is identical to the one described above, except that it uses the (possibly shorter) labels in the set $T'=\{1, 2, ... , \psi(u, \sigma)\}$ to name the $\sigma$-transitions that enter some positive-probability state $u$. By exactly the same argumentation that led to Inequality (\ref{eq:chili}), we get
\begin{equation}\label{eq:psili}
    \mathbb{E}\left[{\sum_{i=1}^n\log_2(\psi(u_i,\sigma_i))}\right]
    \geq n.
\end{equation}

We wish to calculate 
the decrease in the expected length of our coding, i.e. 
\begin{align*}
    \mathbb{E}\left[\codinglength\right] - \mathbb{E}\left[\codinglengthpsi\right].
\end{align*}
Use the linearity of the expectation to obtain
\begin{align*}
    \sum_{i=1}^n\mathbb{E}[\log_2(\chi(u_i,\sigma_i))-\log_2(\psi(u_i,\sigma_i))]=   n\mathbb{E}[\log_2(\chi(u_1,\sigma_1))-\log_2(\psi(u_1,\sigma_1))],
\end{align*}
which follows from the fact that the $u_i$ have the same distribution for all $i$, and the same thing is  true for the $\sigma_i$ as well.

We now use the state $q'$ to obtain a lower bound on this expectation.
\begin{align*}
    \mathbb{E}[\log_2(\chi(u_1,\sigma_1))&-\log_2(\psi(u_1,\sigma_1))]\\
    &= \sum_{(u,\sigma)}\mathbb{P}(u_1 = u, \sigma_1 = \sigma)(\log_2(\chi(u, \sigma)) - \log_2(\psi(u, \sigma))),
\end{align*}
where the sum is taken over state-symbol pairs ($u,\sigma$) which have non-zero probability under the distribution of $(u_1,\sigma_1)$. 
Keeping only the term corresponding to the pair $(q', \gamma)$ on the right-hand side, we get
\begin{align*}
    \mathbb{E}[\log_2(\chi(u_1,\sigma_1))&-\log_2(\psi(u_1,\sigma_1))]\\ 
    &\geq \mathbb{P}(u_1 = q', \sigma_1 = \gamma)(\log_2(\chi(q', \gamma)) - \log_2(\psi(q', \gamma))).
\end{align*}
Since $u_1$ and $\sigma_1$ are independent,  
$\mathbb{P}(u_1 = q', \sigma_1 = \gamma)=\mathbb{P}(u_1=q')\mathbb{P}(\sigma_1=\gamma)$. Both of these probabilities are non-zero, and neither depends on $n$.

We know 
$\psi(q',\gamma) \leq \chi(q', \gamma)-1$.
Using this, we obtain
\begin{align*}
    \log_2(\chi(q', \gamma)) - \log_2(\psi(q', \gamma)) \geq \log_2(\chi(q', \gamma)) - \log_2(\chi(q', \gamma)-1),
\end{align*}
whose right-hand side decreases as the $\chi$-values increase, so 
\begin{align*}
    \log_2(\chi(q', \gamma)) - \log_2(\chi(q', \gamma)-1) \geq
    \log_2(\tau) - \log_2(\tau-1) >0.
\end{align*}
We have established
\begin{align*}
    \mathbb{E}[\log_2(\chi(u_1,\sigma_1))-\log_2(\psi(u_1,\sigma_1))] \geq J > 0,
\end{align*}
where $J$ does not depend on $n$.
This lets us conclude that
\begin{equation}\label{eq:fark}
    \mathbb{E}\left[\codinglength\right] - \mathbb{E}\left[\codinglengthpsi\right] \geq nJ.
\end{equation}
Combining Inequalities (\ref{eq:psili}) and (\ref{eq:fark}), we obtain
\begin{equation}\label{eq:erdos}
\mathbb{E}\left[\codinglength\right] \geq n + nJ 
\end{equation}
for some positive constant $J$ independent of $n$. 
The left-hand side of Inequality (\ref{eq:erdos})  is the expectation of the number of bits $M$ will forget on an  $n$-step subcomputation selected from the described distribution.
By the probabilistic method, there exists a particular $n$-step subcomputation starting from some positive-probability state $u_0$ and consuming a string $\mathbf{s}$  
such that 
\begin{equation*} 
    \codinglength \geq n+nJ.
\end{equation*}
Every state with a positive probability in $D$ is reachable from the initial state $q_0$ of $M$. Let $K$ be the length of the longest simple path that connects $q_0$ to such a positive-probability state, and let $x$ be an input string 
bringing $M$ from $q_0$ to $u_0$ through a simple path. 
The input string $x\mathbf{s}$ (of length at most $n+K$) will force $M$ to forget at least $n+nJ$ bits. This makes at least
\begin{align}\label{fractional end equation}
    \frac{n(1+J)}{n+K} = \frac{1+J}{1 + K/n}
\end{align}
forgotten bits per step. 
For all $n\geq \frac{2K}{J}$,  the value  (\ref{fractional end equation}) is at least $\frac{1+J}{1+J/2}$, and the input string $w=x\textbf{s}$ described above 
makes the machine forget at least
\begin{align*}
    \left(1 + \frac{J/2}{1+J/2}\right)|w|
\end{align*}
bits.
\end{proof}


\section{Concluding remarks}\label{sec:conc}

The approach we present for the study of energy complexity can be extended to several other scenarios, like interactive proof systems, involving finite-memory machines. 
We end with a list of our plans for future research, and open questions.
\begin{itemize}
   \item As we noted, the general QFA model is not flexible enough to study the energy costs of computational steps which forget less information than the maximum amount allowed by the number of operation elements. The formulation of a new model that is able to explicitly represent distinctions between the ``cheap'' and ``costly'' steps of a quantum finite automaton would be helpful in answering further energy complexity questions about those machines.
   \item Are there languages whose bounded-error recognition by QFAs have lower energy complexity than their zero-error recognition? Would the results of Theorems \ref{th:atleastnplus1}, \ref{th:just2} and \ref{th:error} about maximum step costs be helpful in answering this question?
\end{itemize}

\section*{Acknowledgements}

We thank \"{O}yk\"{u} Y\i{}lmaz
and
Meri\c{c} \"{U}ng\"{o}r, who collaborated with us in an earlier stage of this study regarding zero-error QFAs. We are grateful to Utkan Gezer for his invaluable technical assistance. This research was partially supported by Boğaziçi
University Research Fund Grant Number 22A01P1.

\begingroup
\raggedright
\bibliographystyle{splncs04}
\bibliography{references} 
\endgroup

\end{document}